\documentclass[preprint,aps,showpacs,pre,amsmath,amssymb]{revtex4}

\usepackage{amssymb}
\usepackage{graphicx}
\usepackage{dsfont}
\usepackage{color,soul}
\usepackage{xcolor}
\usepackage{appendix}
\usepackage{physics}
\usepackage{enumitem}

\begin{document}

\title{Klein tunneling degradation and enhanced Fabry-Pérot interference in graphene/\textit{h}-BN moiré-superlattice devices}
	
\author{Viet-Anh Tran} \email{anh.tran@uclouvain.be}
\author{Viet-Hung Nguyen} \email{viet-hung.nguyen@uclouvain.be}
\author{Jean-Christophe Charlier} \email{jean-christophe.charlier@uclouvain.be}
\affiliation{Institute of Condensed Matter and Nanosciences, Universit\'{e} catholique de Louvain (UCLouvain), Chemin des \'etoiles 8, B-1348 Louvain-la-Neuve, Belgium}

\begin{abstract}
Hexagonal boron-nitride (\textit{h}-BN) provides an ideal substrate for supporting graphene devices to achieve fascinating transport properties, such as Klein tunneling, electron optics and other novel quantum transport phenomena. However, depositing graphene on \textit{h}-BN creates moiré superlattices, whose electronic properties can be significantly manipulated by controlling the lattice alignment between layers. In this work, the effects of these  moiré structures on the transport properties of graphene are investigated using atomistic simulations. At large misalignment angles (leading to small moiré cells), the transport properties (most remarkably, Klein tunneling) of pristine graphene devices are conserved. On the other hand, in the nearly aligned cases, the moiré interaction induces stronger effects, significantly affecting electron transport in graphene. In particular, Klein tunneling is significantly degraded. In contrast, strong Fabry-Pérot interference (accordingly, strong quantum confinement) effects and non-linear I-V characteristics are observed. P-N interface smoothness engineering is also considered, suggesting as a potential way to improve these transport features in graphene/\textit{h}-BN devices.
\end{abstract}

\maketitle
	
\section{Introduction}

Since its discovery in 2004 \cite{novoselov2004electric}, graphene has become the subject of intensive research and was suggested as candidate for several potential applications, thanks to its unique electronic properties \cite{Novoselov2012,Ferrari2015,Neto2009}. In particular, monolayer graphene exhibits a linear dispersion relation at the six corners  (i.e., K- and K'- points) of its hexagonal Brillouin zone \cite{Neto2009}. These Dirac cones result in massless Dirac fermions with relativistic-like behaviors near the Fermi level and interesting ambipolar characteristics \cite{novoselov2005two}. On this basis, several peculiar phenomena, such as Klein tunneling \cite{Katsnelson2006,Young2009,Pereira2010,du2018tuning,Allain2011,Nguyen2018}, electron optics \cite{Vadim2007,Lee2015,Chen2016,Nguyen2016}, unusual half-integer quantum Hall effect \cite{Bolotin2009,Mark2011}, valleytronics \cite{Rycerz2007,Mrudul2021}, etc. have been explored in graphene. In addition, pristine graphene exhibits exceptionally high mobility of charge carriers \cite{Bolotin2008,Banszerus2015,Du2008,Mayorov2011}, which is an important ingredient for applications in nanoelectronics \cite{Schwierz2013,Dauber2015,Zhong2020,David2021}. Remarkably, the high carrier mobility of graphene makes it an ideal platform for observing ballistic transport phenomena and exploring novel electronic components of Dirac fermions optics \cite{Lee2015,Chen2016,Rickhaus2013,Wilmart2014,Taychatanapat2013,Rickhaus2015,Liu2017,Graef2019,Brun2019,Brun2020,Brun2022}.

Among the transport phenomena reported in graphene, Klein tunneling and Fabry-Pérot interference are two especially fascinating ones. Klein tunneling \cite{Klein1929}, predicted for massless relativistic particles, refers to the feature that electrons can be transmitted through potential barriers without any backscattering. This phenomenon has been indeed observed in monolayer graphene devices, since their charge carriers really exhibit massless Dirac characteristics \cite{Katsnelson2006}. In fact, the perfect matching of electron and hole wave-functions at the barrier interface enables a unit transmission probability in the normal incident direction. Fabry-Pérot interference \cite{Hernandez1988} has been originally explored for light propagation through an optical cavity and obtained when circulating light in the cavity and incident light are in phase. Due to their high mobility, ballistic Dirac fermions in graphene behave in close analogy to light rays in optical media, manifesting in a variety of interference and diffraction effects \cite{Vadim2007,Rickhaus2015,Lee2015,Chen2016,Darancet2009}. Fabry-Pérot interference has been hence also explored in graphene P-N heterostructures \cite{Young2009,Brun2020}. It is worth noting that strong Fabry-Pérot resonances require resistive P-N interfaces, whereas Klein tunneling makes these interfaces more transparent. The latter actually limits Fabry-Pérot resonances, which is a manifestation of the weak confinement of electrons by electrostatic potentials in graphene \cite{Brun2020}.

Due to its 2D nature, the carrier mobility in graphene is however very sensitive to the quality of its supporting substrate. Indeed, scattering is induced by charged surface states and impurities, by substrate surface roughness, by surface optical phonons, etc \cite{Rakheja2013}. In such context, hexagonal boron nitride (\textit{h}-BN) has emerged as one of the best substrates to improve the carrier mobility in graphene \cite{Yin2016}. Owing to the strong, in-plane bonding of the planar hexagonal lattice structure, \textit{h}-BN is relatively inert and free of dangling bonds or surface charge traps. Indeed, high carrier mobilities of $\sim$ $10^4$ cm$^2$/Vs to $10^5$ cm$^2$/Vs and micrometer-scale ballistic transport length have been demonstrated in encapsulated graphene/\textit{h}-BN \cite{Dean2010,Mayorov2011}. Therefore, \textit{h}-BN substrates have been chosen for graphene devices in several transport experiments, most remarkably, such as Klein tunneling \cite{du2018tuning}, Fabry–Pérot \cite{Handschin2017,Ahmad2019} and Aharonov–Bohm interferences \cite{Dauber2017,Ronen2021}, and electron optics phenomena \cite{Chen2016,Graef2019,Brun2019,Brun2022,Boggild2017,Wang2019}.

However, the assembly of graphene and \textit{h}-BN also induces the presence of a moiré superlattice because of their lattice mismatch, and the moiré wavelength is dependent on the misalignment angle between the two hexagonal lattices \cite{Yankowitz2012,Rebeca2018}. A large moiré structure at small misalignment angles presents significant effects on the electronic structure of graphene. In particular, a finite bandgap opens at the Dirac (charge neutrality) point, and second-generation Dirac cones are also observed at finite energies \cite{Rebeca2018,Moon2014}. These electronic properties result in the observation of exciting phenomena such as Hofstadter butterfly and fractional quantum Hall effects \cite{Dean2013,Serlin2020}, valley Hall effect \cite{Arrighi2022}, correlated states \cite{Sun2021}, orbital ferromagnetism \cite{Chen2020}, and so on.

As \textit{h}-BN has often been used as a high-quality substrate for graphene devices, it naturally gives rise to questions about the effects of the mentioned moiré structure on transport phenomena, such as Klein tunneling and Fabry-Pérot interference. Using atomistic simulations, we found that at large misalignment angles (when small moiré structures are induced), the transport properties reported in pristine graphene are preserved in graphene/\textit{h}-BN devices. However, at small misalignment angles, large moiré structure induces significant effects on the electronic transport. Most remarkably, Klein tunneling is found to be significantly degraded whereas strong Fabry-Pérot resonances and strong quantum confinement can be obtained in the nearly aligned cases.

\section{Methodology}

\begin{figure} [h!]
    \centering
    \includegraphics[width=15cm]{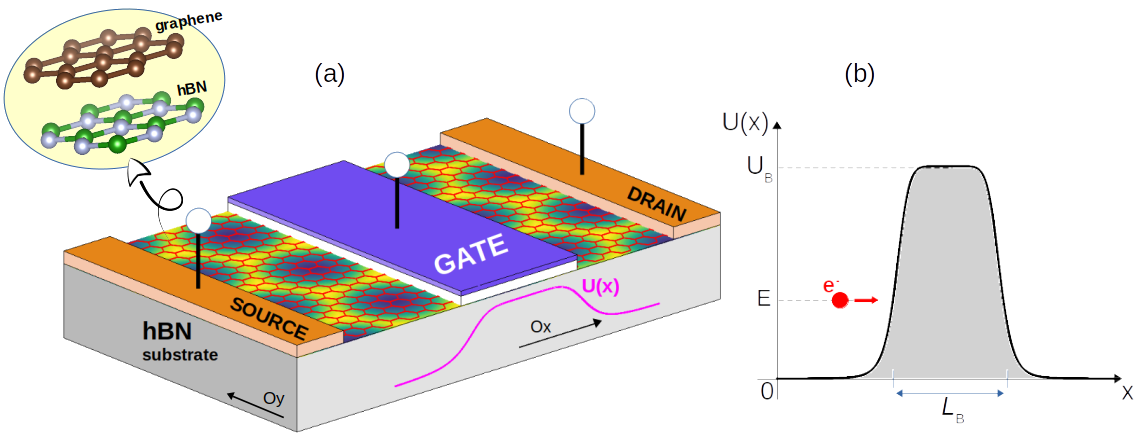}
    \caption{(a) Schematic of a typical device of graphene on \textit{h}-BN substrate. Electronic transmission takes place along the \textit{Ox} axis and gate voltage induces a 1D potential barrier \textit{U(x)}. The inset illustrates the atomic model of graphene coupled to a single \textit{h}-BN layer. (b) Electron transport through the potential barrier $U(x)$ with the barrier height $U_B$ and width $L_B$.}
    \label{fig:figure1}
\end{figure}

Electronic transport through a single potential barrier in graphene on \textit{h}-BN devices is investigated, as schematized in Fig.\ref{fig:figure1}. The transport takes place along the (\textit{Ox}) direction, while a $1D$ potential $U(x)$ is induced and controlled by the gate voltage. The size of graphene sample in the lateral (\textit{Oy}) direction $W_y$ is assumed to be much larger than the gate length (in particular, the barrier width $L_B$ here). In such condition, the graphene channel can be approximately modelled as an infinite and periodic lattice along the \textit{Oy} axis. Since \textit{h}-BN substrate is a large-gap semiconductor, its contribution to the electronic transport of the device is  negligible. On this basis, our atomistic simulations consider a model with graphene coupled to a single \textit{h}-BN layer to compute the effects of moiré structure.

Before computing the electronic transport in the above-described device, graphene/\textit{h}-BN superlattices were first relaxed using molecular dynamics simulations with classical potentials. In particular, intralayer forces are computed using the optimized Tersoff and Brenner potentials \cite{Lindsay2010}, whereas interlayer forces are modeled using the Kolmogorov–Crespi potentials \cite{Leven2016}. The atomic structure was optimized until all the force components were smaller than $0.5$ $meV$/atom. The electronic properties and transport were then computed using the $p_z$ tight-binding Hamiltonian, similar to those in Refs. \cite{Trambly2010,Moon2014},
\begin{eqnarray}
    H &=& H_0 + U(x) \label{eq:hamiltonian}\\
    H_0 &=& \sum_n \epsilon_n a_n^{\dag} a_n +\sum_{n,m} t({\vec r}_{nm}) a_n^{\dag} a_m + h.c \nonumber
\end{eqnarray}
While the tight-binding Hamiltonian $H_0$ computes the electronic properties of graphene/\textit{h}-BN superlattices, $U(x)$ models the gated-induced potential barrier as mentioned above. $\epsilon_n$ represents the onsite energies of atoms, in particular, $\epsilon_n$ = $0$ eV, $3.34$ eV and $-1.4$ eV for carbon, boron and nitrogen atoms, respectively. The hopping energy $t({\vec r}_{nm})$ between n$^{th}$ and m$^{th}$ sites is determined by the Slater-Koster formula
\begin{eqnarray}
t({\vec r}_{nm}) &=& V_{pp\pi}(r_{nm}) \sin^2\phi_{nm} + V_{pp\sigma}(r_{nm})\cos^2\phi_{nm} \\
V_{pp\pi}(r_{nm}) &=& V_{pp\pi}^0\exp\left ( \frac{a_0 - r_{nm}}{r_0} \right ) \nonumber \\
V_{pp\sigma}(r_{nm}) &=& V_{pp\sigma}^0\exp\left ( \frac{d_0 - r_{nm}}{r_0} \right ) \nonumber
\end{eqnarray}
where the direction cosine function of ${\vec r}_{nm}$ along Oz axis $\cos \phi_{nm} = z_{nm}/r_{nm}$, $r_0 = 0.184a$, $a_0 = a/\sqrt 3$ with $a \simeq 2.49$\AA ~ and $d_0 = 3.35$\AA. As $U$ is a function of $x$ only, the Hamiltonian \eqref{eq:hamiltonian} satisfies the translational periodicity in the $Oy$ direction and can therefore be rewritten in a wavevector $k_y$-dependent (quasi-$1D$) form $H(k_y)$. In addition, the potential barrier is practically a smooth function of $x$, and its smoothness depends on the thickness as well as dielectric constant of the gate-insulating layer. To mimic this property, the potential $U(x) = U_B / [1 + (2x/L_B)^\alpha]$ \cite{Brun2020} was used in our simulations, with barrier height $U_B$ and width $L_B$ (see Fig.1(b)). Here, $\alpha$ is the smoothness parameter. In particular, the larger it is, the more abrupt the barrier is and the ideally  abrupt one corresponds to $\alpha = \infty$.

The Hamiltonian \eqref{eq:hamiltonian} was solved using Green's function technique \cite{Nguyen2023}. In particular, the retarded Green's function is computed as
\begin{equation}
    \mathcal{G}(E,k_y) = [E+i0^+-H_{device}(k_y)-\Sigma_L(E,k_y) - \Sigma_R(E,k_y)]^{-1} 
\end{equation}
where $\Sigma_{L,R}(E,k_y)$ are self-energies representing the left and right lead-to-device couplings, respectively. The transmission function is determined as $\mathbb{T}(E,k_y) = \mathrm{Tr} (\Gamma_L \mathcal{G} \Gamma_R \mathcal{G}^\dag)$ with $\Gamma_{L,R} = i(\Sigma_{L,R} - \Sigma^{\dag}_{L,R})$. The current and low-bias conductance are then estimated using the Landauer formula:
\begin{eqnarray}
J(V_b) &=& \frac{e W_y}{\pi h} \int_{BZ} dk_y \int dE{\color{white},}\mathbb{T}(E,k_y){\color{white},}\left [ f_L(E) - f_R(E)  \right ], \\
G(E_F)  &=& \frac{e^2 W_y}{\pi h} \int_{BZ} dk_y \int dE{\color{white},}\mathbb{T}(E,k_y){\color{white},}\left [ -\frac{\partial f(E)}{\partial E}  \right ],
\end{eqnarray}
where $f_{L,R} (E) = \left [ 1 + \exp \left ( \frac{E - E_{FL,FR}}{k_b T} \right ) \right ]^{-1}$ are the Fermi distribution functions in the left and right leads, respectively, with the Fermi levels $E_{FL,FR}$ and $E_{FR} = E_{FL} - eV_b$. In addition, the local density of states (LDOS) can also be computed using Green's function: $D(E,k_y,r_n) = - \Im[G_{n,n}(E,k_y)]/\pi$. Finally, note that as the number of atoms in the supercell of nearly-aligned graphene/\textit{h}-BN is huge (particularly, 12322 atoms for $\theta = 0^\circ$), making standard calculations unfeasible, the Green's function equation (3) is solved using the recursive techniques developed in Ref. \cite{Nguyen2023}.

\section{Results and discussions} 

\begin{figure} [t!]
    \centering
    \includegraphics[width=14cm]{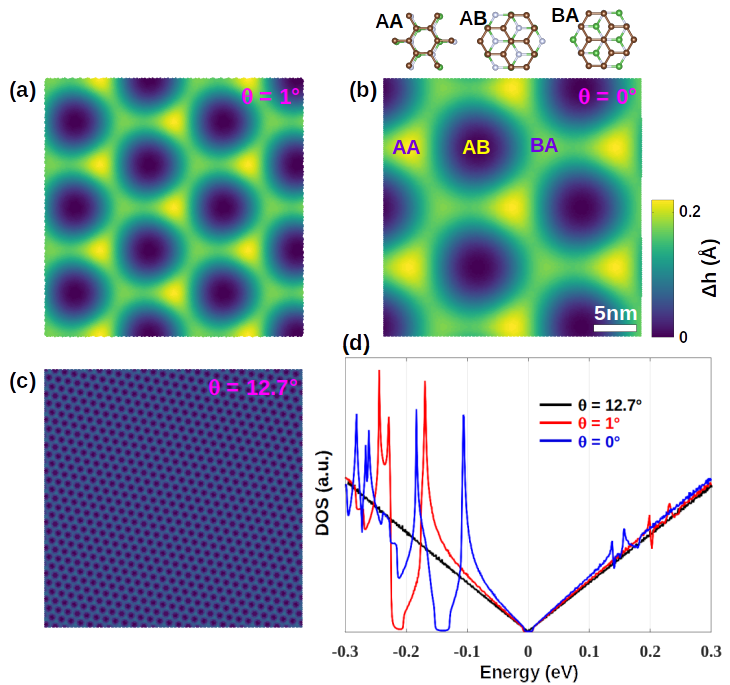}
    \caption{Moiré atomic structures (a-c) and densities of electronic states (d) of largely misaligned ($\theta = 12.7^\circ$) and nearly aligned ($\theta = 1^\circ$ and $0^\circ$) graphene on \textit{h}-BN. Local stacking configurations within the obtained moiré cell are depicted in (b). Buckling is measured by the spatial variation of height $\Delta$h(\AA) of the graphene sheet along Oz axis.}
    \label{fig:figure2}
\end{figure}

First, the atomic structure of the moiré superlattice is discussed. Figs.2(a-c) visualize the buckling of the moiré structures obtained in graphene by MD simulations when varying the misalignment angle $\theta$ with the \textit{h}-BN layer. These results are shown to be in good agreement with the reported STM images \cite{Yankowitz2012}. In particular, at large $\theta$-angles, moiré structures have a small wavelength and present negligible effects on graphene. Indeed, the graphene sheet is almost flat as illustrated in Fig.\ref{fig:figure2}(c) for $\theta \simeq 12.7^\circ$. When decreasing $\theta$, especially in the nearly aligned cases, the moiré structures become more visible and display a significantly large wavelength, i.e., about $10$ nm at $\theta = 1^\circ$ (Fig.\ref{fig:figure2}(a)) and $14$ nm at $\theta = 0^\circ$ (Fig.\ref{fig:figure2}(b)). 
Consequently, the electronic structure of free-standing graphene is preserved at large $\theta$-angles, i.e., the V-shaped density of states (DOS) signifying gapless linear dispersion (see also in the Supplementary Material (SM)) is still observed, as illustrated in Fig.\ref{fig:figure2}(d). At smaller angles, the large moiré structures significantly affect the electronic properties of graphene. In particular, a moiré induced-bandgap at zero energy ($\sim$ $15$ meV for $\theta = 0^\circ$), second-generation Dirac points and spatially dependent electronic properties are observed (see Fig.\ref{fig:figure2}(d) and SM - section A). These electronic DOS are in good agreement with experimental measurements \cite{Yankowitz2012,Hakseong2018,Rebeca2018}.
Note that a negligible bandgap is also obtained in Ref. \cite{Moon2014} by similar calculations performed in the unrelaxed lattices. This emphasizes the importance of atomic relaxation effects on the electronic properties of nearly aligned graphene/\textit{h}-BN superlattices.

\begin{figure} [t!]
    \centering
    \includegraphics[width=10.5cm]{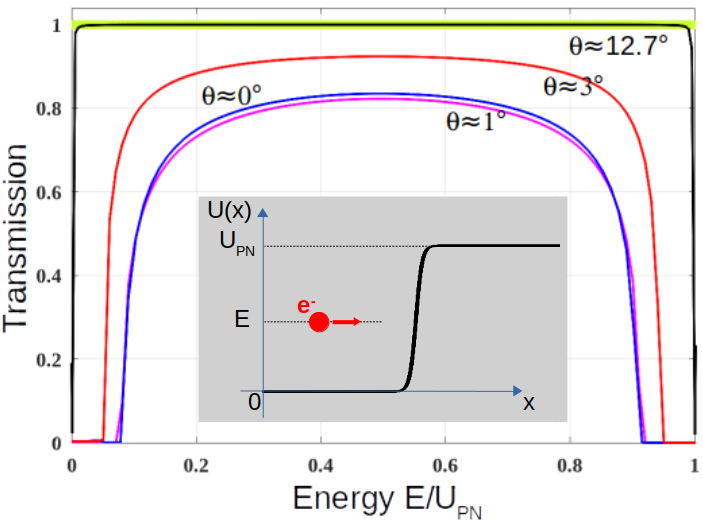}
    \caption{Electronic transport through a single P-N junction in graphene/\textit{h}-BN devices with different misalignment angles $\theta$: $0^\circ$, $1^\circ$, $3^\circ$, and $12.7^\circ$. The transmission function is estimated along the normal incident direction ($\phi = 0^\circ$, parallel to the Ox axis). Green line illustrates perfect transmission obtained in pristine graphene devices.}
    \label{fig:figure3}
\end{figure} 

As a potential barrier (see Fig.1) consists of two parallel P-N junctions, electronic transport through a single P-N junction in graphene on \textit{h}-BN devices is first investigated. In Fig.\ref{fig:figure3}, the electronic transmission in the normal incident direction ($\phi = 0^\circ$) is computed and presented for graphene/\textit{h}-BN P-N junctions with four different misalignment angles. It is worth noting that because of its Dirac conical electronic structure, this transmission in pristine graphene devices also takes place without any backscattering ($\mathbb{T}(E,\phi = 0^\circ) = 1$ $\forall \, E$) \cite{Allain2011}. This implies that the pristine graphene P-N junctions are perfectly transparent for $\phi = 0^\circ$, manifesting the Klein tunneling effect. In the graphene/\textit{h}-BN cases, these transport properties are conserved at large misalignment angles, as demonstrated by the results for $\theta = 12.7^\circ$ in Fig.\ref{fig:figure3}. This result is consistent with the negligible effects of moiré structure at such angle, as discussed in Fig.\ref{fig:figure2}. However, when decreasing $\theta$, a significant reduction in the transmission function is observed in the range $E \in [0,U_{PN}]$, implying that these graphene/\textit{h}-BN P-N junctions are no longer perfectly transparent. This result is a direct consequence of the above-discussed effects of large moiré structure on the electronic properties of graphene (see Fig.\ref{fig:figure2}(d)).

\begin{figure} [h!]
    \centering
    \includegraphics[width=16.5cm]{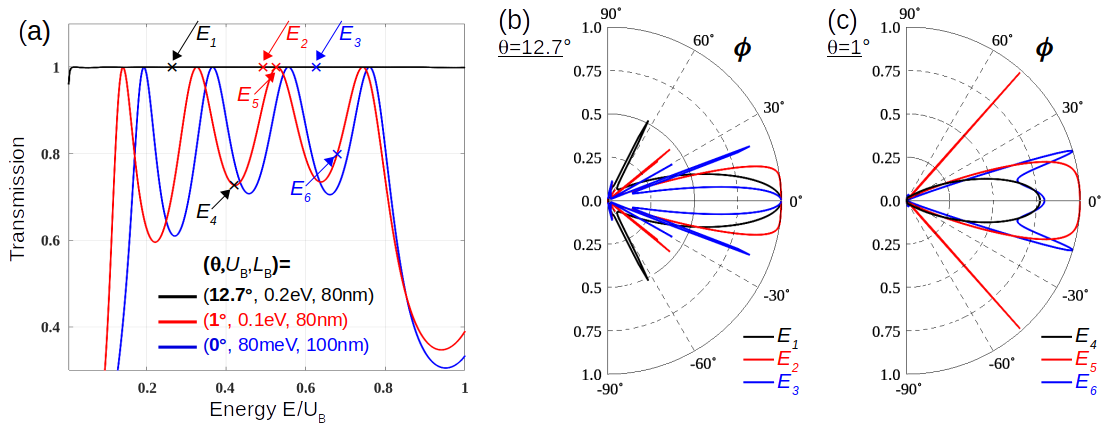}
    \caption{Electronic transport through a single potential barrier in graphene/\textit{h}-BN devices.  Transmission function in the normal directionb ($\phi = 0^\circ$) is presented for different misalignment angles $\theta$ in (a). The smoothness parameter of the barrier is $\alpha = 8$. The other barrier parameters, $U_B$ and $L_B$, are mentioned in the frame. Transmission function versus incident direction ($\phi \in [-90^\circ,90^\circ]$) is presented for $\theta = 12.7^\circ$ in (b) and for $\theta = 1^\circ$ in (c) at different energies marked in (a).}
    \label{fig:figure4}
\end{figure}

The electronic transport properties through a potential barrier in the graphene/\textit{h}-BN devices are computed in Fig.\ref{fig:figure4}. First, the transmission function in the normal incident direction is presented in Fig.\ref{fig:figure4}(a) for the three misalignment angles. Note again that in pristine graphene devices, this normal-incident transmission represents the Klein tunneling effect, which has been explained by the conservation of the sublattice pseudo-spin of massless Dirac fermions \cite{Katsnelson2006,Allain2011}. Indeed, such a feature is preserved in graphene/\textit{h}-BN devices with large misalignment angles, as demonstrated for $\theta = 12.7^\circ$ in Fig.\ref{fig:figure4}(a). This result is actually consistent with the transport properties through the P-N junctions, as discussed above. Indeed, \textit{h}-BN layers simply act as a flat substrate, thus not affecting the conservation of sublattice pseudo-spin. However, when decreasing $\theta$, thus increasing the moiré wavelength, the considered transmission is overall reduced and some resonant peaks are observed. Indeed, the transmission is not always perfectly equal to $1$ in the energy 
range $E \in [0,U_{B}]$, as obtained in the cases of large $\theta$. Actually, electrons described by the 2D massless Dirac equation in graphene carry a pseudo-spin related to their freedom of belonging either to sublattice A or to sublattice B. In pristine graphene, sublattice symmetry is present, and accordingly, the conservation of the sublattice pseudo-spin can be obtained in the normal incident transmission through potential barriers, thus resulting in the observation of Klein tunneling. These features are conserved in largely misaligned graphene/\textit{h}-BN systems, because \textit{h}-BN acts as a flat substrate (when moiré effects are negligible) and therefore does not affect the sublattice symmetry and, accordingly, does not affect the conservation of sublattice pseudo-spin in graphene. The picture however changes drastically when the large moiré structure is obtained at small misalignment angles. In particular, different (i.e., AA, AB, and BA) stacking and large regions are formed within the moiré cell, essentially due to the atomic reconstruction effects in small-$\theta$ superlattices, as illustrated in Fig.\ref{fig:figure2}(b). In these stacking regions, the electronic couplings with the B and N atoms of the substrate locally break the sublattice symmetry of graphene, as illustrated by the LDOS images projected on its two sublattices (see SM, Fig.S2). On the one hand, these sublattice symmetry breakings lead to a finite bandgap opening as discussed above. More importantly, because of the discussed sublattice symmetry breaking, the conservation of the sublattice pseudo-spin can no longer be satisfied in the transmission through potential barriers (see SM - section B for more details), thus essentially explaining the observed degradation of Klein tunneling.

Figs.3 and 4(a) additionally suggest that although perfect transmission (Klein tunneling) is overall no longer observed in large moiré superlattices, P-N junctions and potential barrier devices exhibit different behaviors. In particular, resonant peaks occur for potential barriers, whereas the transmission is reduced over the whole energy range computed for P-N junctions. Interestingly, this result is in good agreement with the principles of Fabry-Pérot interference, implying an optical-like behavior of electrons in graphene. In particular, resistive P-N interfaces are the key ingredient for generating circulating electron waves inside the potential barriers, leading to Fabry-Pérot resonance when the incident wave aligns in phase with these circulating ones. In pristine graphene and largely misaligned graphene/\textit{h}-BN devices, the presence of Klein tunneling in the nearly normal incident directions makes P-N interfaces highly transparent. Therefore, 
Fabry-Pérot interference can only observed in these devices in the very oblique directions $\phi$ \(\gg\) $\phi_0 \equiv 1/\sqrt{\pi k_F d}$ \cite{Allain2011}. Here $\phi_0$ is the collimation angle determined with the Fermi momentum $k_F$ and the barrier smoothness length $d$. However, these properties are drastically changed in nearly aligned graphene/\textit{h}-BN devices as P-N interfaces become resistive and hence significant Fabry-Pérot interference can be observed even in the normal incident direction. For completeness, Figs.4(b-c) present the angular-dependent transmission computed at the energies marked in Fig.\ref{fig:figure4}(a) and for two angles $\theta = 12.7^\circ$ and $1^\circ$. In a short summary, together with the data depicted in Fig.\ref{fig:figure4}(a), the presented results emphasize that
    (1) at large misalignment angles, Klein tunneling is observed in the nearly normal direction whereas Fabry-Pérot interference occurs in the very oblique ones;
    (2) in nearly aligned cases, as the Klein tunneling is strongly degraded, significant Fabry-Pérot resonances can be obtained in all incident directions.
\begin{figure} [h!]
    \centering
    \includegraphics[width=15.5cm]{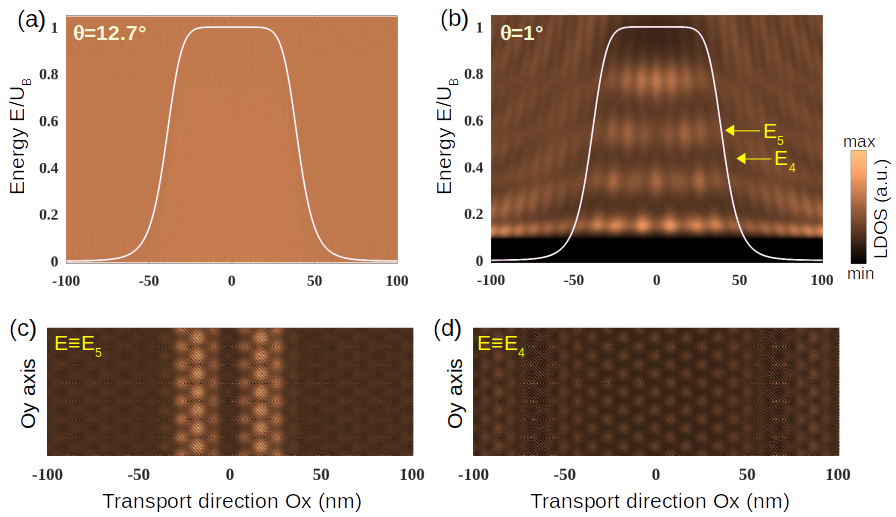}
    \caption{Local densities of states computed for electrons propagating in the normal direction as a function of energy in (a,b) with the barrier parameters ($\theta$,$U_B$,$\L_B$) as given in Fig.\ref{fig:figure4}(a). 2D maps of LDOS computed for $\theta = 1^\circ$ at the resonant ($E = E_5$, see also in Fig.4(a)) and off-resonant ($E = E_4$, see also in Fig.4(a)) states depicted in (b) are illustrated in (c,d), respectively.}
    \label{fig:figure5}
\end{figure}

The above-discussed transparency and resistivity of the P-N interfaces and related transport properties at large and small misalignment angles are visibly illustrated by the LDOS in Figs.5(a-b). In addition, it is well known that the presence of Klein tunneling is an obstacle to graphene-based systems that prevents the confinement of electrons by electrostatic potentials. This obstacle similarly occurs in largely misaligned graphene/\textit{h}-BN devices, as illustrated in Fig.\ref{fig:figure5}(a). In particular, the potential barrier is perfectly transparent, which is consistent with the perfect transmission of the electron wave discussed above. In contrast, relatively strong confinements of electrons is observed in the nearly aligned graphene/\textit{h}-BN case (see Fig.\ref{fig:figure5}(b)), which is in accordance with the observation of the Fabry-Pérot resonance in Fig.\ref{fig:figure4}(a). Figs.5(c-d) present 2D maps of the LDOS obtained at two energy levels, resonant ($E_R$) and off-resonant ($E_{OR}$) states, in the device with $\theta = 1^\circ$ in Fig.\ref{fig:figure5}(b). These images clearly characterize the signature of moiré structure, which governs all the transport properties discussed above.
\begin{figure} [h!]
    \centering
    \includegraphics[width=15cm]{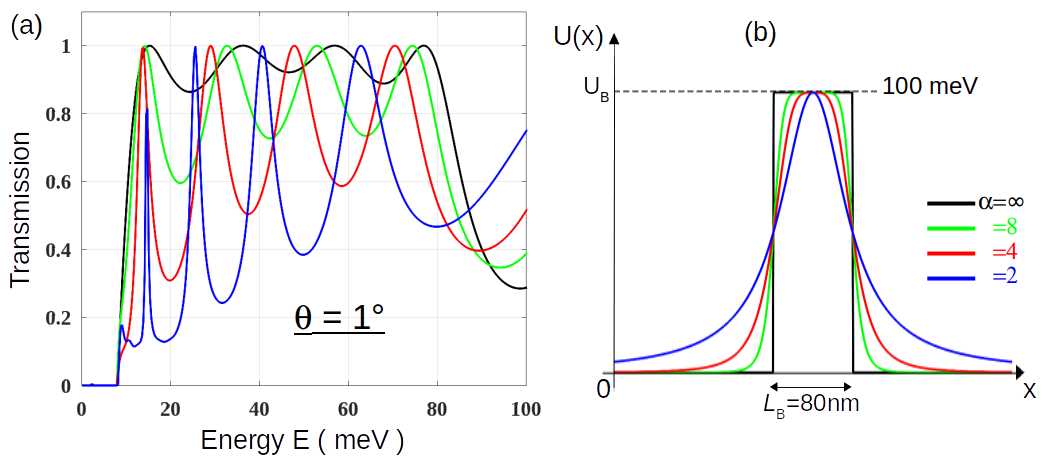}
    \caption{(a) Electronic transmission through potential barriers in the normal direction with different smoothness parameters in the nearly aligned ($\theta = 1^\circ$) graphene/\textit{h}-BN device. Barrier parameters are defined on the right (b).}
    \label{fig:figure6}
\end{figure}

The electronic transport in graphene devices has been demonstrated to be also strongly dependent on the smoothness of both P-N junctions \cite{Cheianov2006} and potential barriers \cite{Brun2020}. In principle, while the barrier height and width are determined and controlled by gate voltage and gate length, respectively, the barrier shape (i.e., smoothness) is dependent on the gate insulator layer, in particular, on its thickness and its dielectric constant. While varying the barrier height and width mainly leads to a change in the number of resonant peaks \cite{Nam2008}, barrier smoothness engineering has been suggested as a way to enhance the Fabry–Pérot effect in graphene \cite{Brun2020}. A larger enhancement is even observed in nearly aligned graphene/\textit{h}-BN devices, as depicted in Fig.\ref{fig:figure6}, compared to the effect in pristine graphene and largely misaligned graphene/\textit{h}-BN ones. In particular, the Klein tunneling in the nearly normal incident directions is unaffected by the mentioned smoothness engineering in the latter devices. However, the Fabry–Pérot effect in the normal incident direction is significantly enhanced when increasing the barrier smoothness (i.e., reducing $\alpha$) in nearly aligned graphene/\textit{h}-BN devices as illustrated for $\theta = 1^\circ$ in Fig.6. Indeed, the variation of transmission $\Delta \mathbb{T} / \mathbb{T}_{peak} $ gets a large value $\gtrapprox 80\%$ for $\alpha = 2$ while it is only about $15\%$ for the abrupt barrier ($\alpha = \infty$). These results also imply that the degradation of Klein tunneling in nearly aligned graphene/\textit{h}-BN devices can be significantly enlarged by barrier smoothness engineering.
\begin{figure} [b!]
    \centering
    \includegraphics[width=16.5cm]{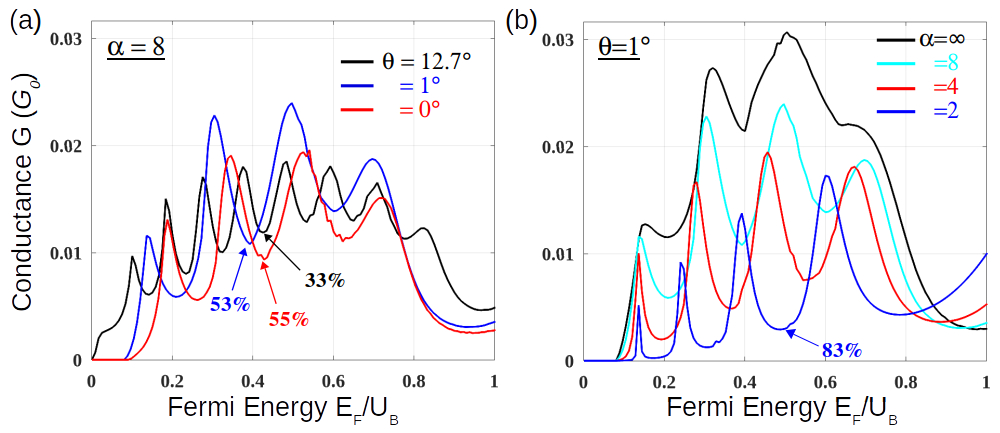}
    \caption{Conductances computed in single barrier graphene/\textit{h}-BN devices with (a) different misalignment angles and (b) different barrier smoothness parameters (for $\theta = 1^\circ$).  Other barrier parameters ($U_B$ and $L_B$) are the same as given in Fig.\ref{fig:figure4}(a). Some estimated values of conductance variation $\mathrm{\Delta G/G_{peak}}$ are indicated by arrows.}
    \label{fig:figure7}
\end{figure}

The electrical conductance, which is a characteristic quantity determined by the contribution of all transmission directions, is also investigated since it is often measured in standard experiments. In Fig.\ref{fig:figure7}(a), the conductance at $0K$ is computed as a function of Fermi energy for three different $\theta$ angles. In the largely misaligned devices, the Fabry–Pérot effect occurs in oblique transmission directions, thus explaining the observed conductance resonant peaks for $\theta = 12.7^\circ$. However, the presence of Klein tunneling in nearly normal incident directions diminishes the magnitude of this resonance, similar to that observed in pristine graphene devices. When decreasing the $\theta$ angle, the degradation of Klein tunneling enlarges the conductance Fabry–Pérot resonance, in particular, the variation of conductance $\mathrm{\Delta G/G_{peak}} \simeq 53\%$ and $55\%$ for $\theta = 1^\circ$ and $0^\circ$, respectively, compared to $33\%$ for $\theta = 12.7^\circ$. In accordance with the effects of barrier smoothness engineering discussed in Fig.\ref{fig:figure6}, much stronger Fabry–Pérot resonances are obtained in the smoother barrier device, as demonstrated in Fig.\ref{fig:figure7}(b). Indeed, $\mathrm{\Delta G/G_{peak}} \simeq 83\%$ is observed for $\alpha = 2$ in the device with $\theta = 1^\circ$. Note that such a large enhancement cannot be achieved for largely misaligned graphene/\textit{h}-BN devices, due to the presence of Klein tunneling that is independent on the shape of potential barriers.
\begin{figure} [b!]
    \centering
    \includegraphics[width=15.5cm]{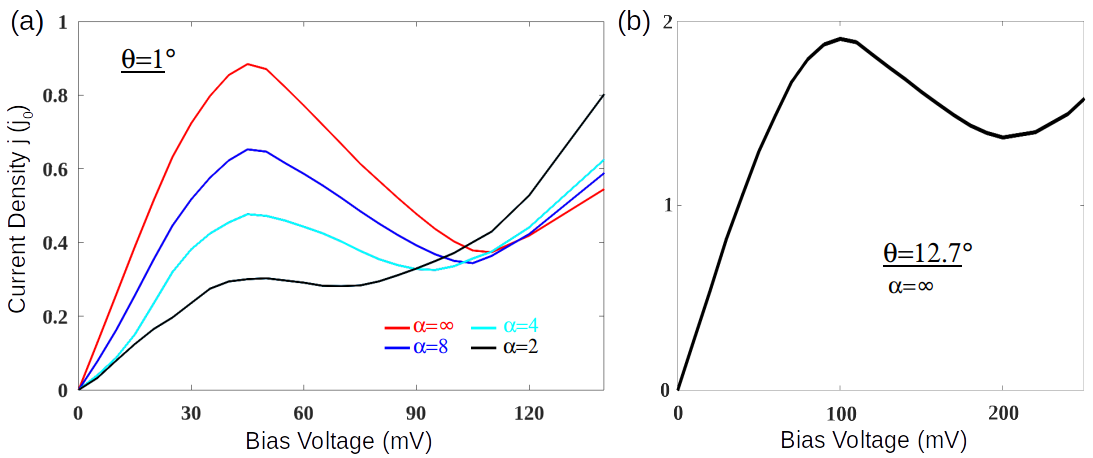}
    \caption{I-V characteristics of the graphene/\textit{h}-BN device of $\theta = 1^\circ$ is computed in (a) for different barrier smoothness parameters. The results in the $\theta = 12.7^\circ$ case is added in (b) for a comparison. Other barrier parameters ($U_B$ and $L_B$) are the same as given in Fig.\ref{fig:figure4}(a).}
    \label{fig:figure8}
\end{figure}

At last, with their high carrier mobility, graphene devices have been demonstrated to be highly efficient for high-frequency applications \cite{Wu2012,Colmiais2022}. These potential applications are extended even when current saturation or negative differential resistance (NDR) occurs \cite{Meric2008,Alarcon2013,Wu2012b,Britnell2013}. Such non-linear characteristics are obtained and essentially explained by the bottleneck effect on the transmission induced by the V-shaped DOS of graphene \cite{Nam2008,Nguyen2015}. The I-V characteristics of the graphene/\textit{h}-BN devices at $0K$ are also computed and presented in Fig.\ref{fig:figure8}. Indeed, the NDR effect can be obtained in both cases of nearly aligned (Fig.8(a)) and misaligned (Fig.8(b)) graphene/\textit{h}-BN. Interestingly, a much larger NDR is obtained in the nearly aligned case, compared to the largely misaligned one. In particular, a peak-to-valley ratio of approximately 2.4 is obtained for $\theta = 1^\circ$ (see $\alpha = \infty$ in Fig.8(a)) while it is only about 1.3 for $\theta = 12.7^\circ$. This enhanced NDR can be essentially explained by the effect of the bandgap opening at $\theta = 1^\circ$. Note that the peak current is obtained at low bias when the tunneling transmission through the barrier presents a large contribution whereas the valley current occurs when the contribution of such tunneling transmission is minimized, thanks to the above-mentioned bottleneck effect at large bias as explained in details in Refs.\cite{Nam2008,Nguyen2015}. When a bandgap opens (in particular, at $\theta = 1^\circ$, compared to the $\theta = 12.7^\circ$ case), such bottleneck effect is enlarged. This significantly reduces the valley current while the peak current is more weakly affected. On this basis, a stronger NDR is achieved at $\theta = 1^\circ$, compared to that obtained at $\theta = 12.7^\circ$. In addition, on contrary with the effects of barrier smoothness engineering  on Fabry-Pérot resonances, the NDR behavior is weakened when the barrier smoothness is enlarged (when desceasing $\alpha$). This is simply due to the reduction of the above-discussed tunneling transmission through the smooth barrier as illustrated in Fig.7(b). This reduction leads to the reduction of current peak as seen in Fig.8(a), thus explaining the weakened NDR with small $\alpha$ as observed.

\section{Conclusion}
Using atomistic simulations, the effects of moiré structure induced by an \textit{h}-BN substrate on the electronic transport properties of graphene devices are investigated. In the largely misaligned cases, \textit{h}-BN simply acts as a flat substrate, and all the electronic and transport properties previously observed in pristine graphene are conserved in the considered device. However, in nearly aligned graphene/\textit{h}-BN superlattices, the large moiré structure significantly alters the electronic properties of graphene and, therefore, its transport properties. In particular, couplings with the B and N atoms break the sublattice symmetry, leading to a bandgap opening in graphene and, on this basis, the significant degradation of Klein tunneling is explored in the electronic transmission through P-N junctions and potential barriers. Instead of having a perfect unity value, transmission through potential barriers in nearly normal incident directions exhibits a strong Fabry–Pérot effect. Accordingly, an enhanced quantum confinement of electrons is observed. Barrier smoothness engineering is also investigated and found to significantly strengthen the observed Fabry–Pérot interference. Consequently, large conductance resonant peaks (i.e., $\mathrm{\Delta G/G_{peak}} > 80\%$) can be obtained in nearly aligned graphene/\textit{h}-BN devices with a smooth barrier. Finally, nonlinear I-V characteristics with enhanced NDR behavior are also predicted in the graphene/\textit{h}-BN large-moiré device. Thus, the present study clarifies the effects of \textit{h}-BN subtrates in graphene devices and suggests that controlling the moiré structure could be a potential way to efficiently engineer their corresponding transport properties.

\section*{Acknowledgments} We acknowledge financial support from the European Union’s Horizon 2020 Research Project and Innovation Program — Graphene Flagship Core3 (N$^{\circ}$ 881603), from the Flag-Era JTC projects “TATTOOS” (N$^{\circ}$ R.8010.19) and “MINERVA” (N$^{\circ}$ R.8006.21), from the Pathfinder project “FLATS” (N$^{\circ}$ 101099139), from the F\'ed\'eration Wallonie-Bruxelles through the ARC project “DREAMS” (N$^{\circ}$ 21/26-116) and  the EOS project “CONNECT” (N$^{\circ}$ 40007563), and from the Belgium F.R.S.-FNRS through the research project (N$^{\circ}$ T.029.22F). Computational resources have been provided by the CISM supercomputing facilities of UCLouvain and the C\'ECI consortium funded by F.R.S.-FNRS of Belgium (N$^{\circ}$ 2.5020.11). V.-H.N. thanks Dr. Xuan-Hoang TRINH for his helps in implementation of numerical codes to compute the lattice atomic structure relaxation.

\bibliographystyle{unsrt}
\bibliography{paper_references_ver4}

\clearpage
\newpage
\renewcommand{\thefigure}{S\arabic{figure}}
\setcounter{page}{1}    
\setcounter{figure}{0}    

\begin{flushleft} 
{\Large \textbf{ \underline {Supplementary materials:}}}
\\
{\color{white} an empty line} \\
{\color{white} an empty line} \\

\textbf{A. Electronic properties of graphene/\textit{h}-BN system} 
\end{flushleft}

In this section, the electronic properties of graphene/\textit{h}-BN moiré superlattices obtained by tight-binding calculations are presented. In particular, the band structures of three misalignment angles ($\theta = 12.7^\circ$, $1^\circ$ and $0^\circ$) are computed in Fig.S1. These results further emphasize the negligible effect of moiré structure in largely misaligned systems ($\theta = 12.7^\circ$) whereas significant effects are obtained in the nearly aligned cases ($\theta = 1^\circ$ and $\theta = 0^\circ$). In particular, at $\theta = 1^\circ$ and $\theta = 0^\circ$, a finite bandgap and the signature of superlattice Dirac points are observed in the obtained bandstructures. Accordingly, the density of states (DOS) and local density of states (LDOS) in these superlattices are also presented in Fig.S2. Remarkably, the LDOS images at the bottom of Fig.S2 illustrate clearly the effects of moiré structure on electronic properties of graphene, leading to the spatial variation as well as the sublattice symmetry breaking in the nearly aligned cases. It is worth emphasizing that these features are not obtained in largely misaligned systems as demonstrated for $\theta = 12.7^\circ$, where all the properties observed in pristine graphene are conserved. 

{\color{white} an empty line} \\

\begin{figure} [h!]
    \centering
    \includegraphics[width=16.5cm]{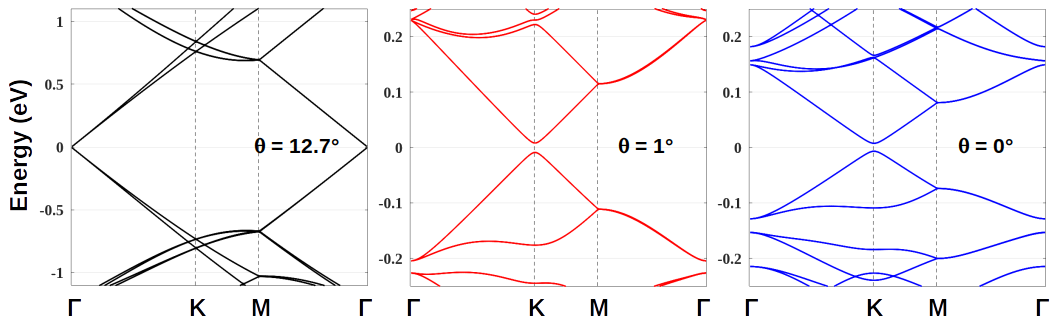}
    \caption{Electronic bandstructures of graphene/\textit{h}-BN moiré superlattice: tight-binding calculations for different misalignment angles $\theta$.}
\end{figure}

{\color{white} an empty line} \\
{\color{white} an empty line} \\
{\color{white} an empty line} \\
{\color{white} an empty line} \\
{\color{white} an empty line} \\

\begin{figure} [h!]
    \centering
    \includegraphics[width=16cm]{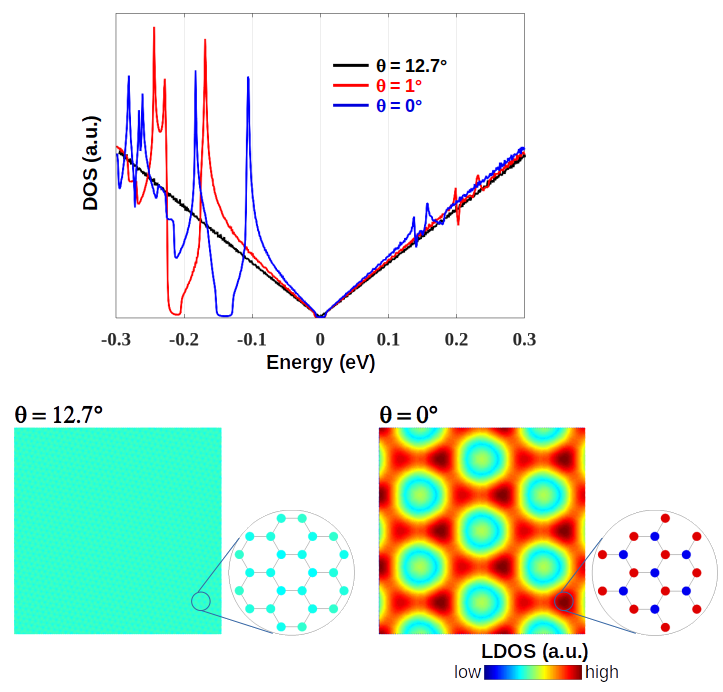}
    \caption{(Top) Densities of states of graphene/\textit{h}-BN moiré superlattices for three misalignment angles $\theta$. (Bottom) Local densities of states at E = 50 meV presents their spatial variation at $\theta = 0^\circ$ (large moiré superlattice), compared to the case of $12.7^\circ$ (small moiré structure). The zoom images of LDOS in bottom illustrate the sublattice symmetry at $\theta = 12.7^\circ$ and the corresponding breaking at $\theta = 0^\circ$.}
\end{figure}
\bigskip
\bigskip
\bigskip
\bigskip
\bigskip

\begin{flushleft}
\textbf{B. Degradation of Klein tunneling with the sublattice symmetry breaking}
\end{flushleft}

Here, a simple demonstration of the degradation of Klein tunneling is presented, when the sublattice symmetry of graphene is broken. To model this problem, the Dirac Hamiltonian in graphene is written as
\begin{equation}
    \widehat{H} = \hbar v_F (\hat \sigma_x k_x + \hat \sigma_y k_y) + \hat \sigma_z \Delta + U(x)
    \nonumber
\end{equation}
with the Pauli matrices $\hat \sigma_{x,y,z}$, the Fermi velocity $v_F$ and a 1D potential energy \textit{U(x)} as investigated in the main text. The term $\Delta$ is added to mimic the sublattice symmetry breaking induced by the electronic couplings between graphene and \textit{h}-BN lattices.

The velocity operator in the \textit{Ox}-direction is defined by
\begin{equation}
    \hat{v}_x = -\frac{i}{\hbar} [\hat{x},\widehat{H}] = v_F \hat{\sigma}_x
    \nonumber
\end{equation}
Its time evolution is given by the Heisenberg equation of motion
\begin{equation}
    \Dot{\hat{v}}_x = - \frac{i}{\hbar} [\hat v_x, \widehat H] = 2 \frac{v_F}{\hbar} (\hbar v_F k_y \hat \sigma_z - \Delta \hat \sigma_y)
    \nonumber
\end{equation}
Note that because of the translational invariance along the \textit{Oy}-direction, the momentum $k_y$ is a conserved quantity as $\dot{k}_y = -i[k_y,\widehat H]/\hbar = 0$, leading to $k_y(t) = k_y(0)$. On this basis, $\Dot{\hat{v}}_x = -2 v_F \Delta \hat \sigma_y / \hbar$ is obtained for $k_y = 0$. 

If $\Delta = 0$, the velocity (or the pseudo-spin) along \textit{Ox} for $k_y = 0$ is a constant of the motion, i.e., $\Dot{\hat{v}}_x = 0$ or $\bra{\psi(t)}\hat v_x \ket{\psi(t)} = \bra{\psi(0)}\hat v_x \ket{\psi(0)}$,  indeed implying the absence of backscattering (observation of Klein tunneling) in pristine graphene \cite{Allain2011} and largely misaligned graphene/\textit{h}-BN devices.

When the sublattice symmetry of graphene is broken (i.e., $\Delta \neq 0$), the above-described properties are no longer obtained, which essentially explains the degradation of Klein tunneling as reported in the main text.

\end{document}